\documentclass{JHEP}
\usepackage{epsfig}

\def\a{\alpha}              \def\g{\gamma}

\def\G{\Gamma}

\newcommand{\ds}{\displaystyle}

\title{What happens to Lattice Fermion near Continuum Limit?}

\author{Minoru Koseki, Hiroto So and Naoya Ukita 
 \\Department of Physics, Niigata University
 \\ Ikarashi 2-8050, Niigata 950-2181, Japan
 \\ E-mail: \email{koseki@muse.hep.sc.niigata-u.ac.jp},
            \email{so@muse.hep.sc.niigata-u.ac.jp},
            \email{ukita@muse.hep.sc.niigata-u.ac.jp}}

\abstract {A Ginsparg-Wilson Relation (GWR) is obtained in the presence
of chiral symmetry breaking terms.  It leads to the PCAC relation as
well as an anomaly relation on the lattice.  For general fermions, the
deviation from the exact GWR is getting small when the block-spin
transformations are performed iteratively.  Based on a simple
geometrical interpretation of the Dirac operator satisfying the GWR, we
find some physical properties shared by the lattice fermions near the
continuum limit.  In two-dimensions, we explicitly construct the GW
Dirac operator by using a conformal mapping.}


\preprint{NIIG-DP-00-2}

\begin{document}

\setcounter{footnote}{0}

\newpage

\begin{flushleft}
{\large{\bf 1.~ Introduction}}
\end{flushleft}

The chiral symmetry on the lattice has been found by L\"uscher \cite{lu}
 based on the Ginsparg-Wilson Relation (GWR) \cite{gw}.  He has opened a
 new era in the study of the regularized theory of chiral fermion.  His
 formalism has been applied to chiral gauge theories, numerical calculations
 of chiral dynamics and supersymmetry on the lattice [3-17].

The purpose of this paper is to investigate the realization of the
chiral symmetry near the continuum limit. The related problem was
discussed by Hasenfratz based on his idea of the perfect action
\cite{hf}: he obtained the relation equivalent to the GWR using the
block-spin transformation(BST).  However, his discussion was restricted
to the behaviors of lattice fermions on the renormalized trajectory.

We first consider a BST for chiral non-invariant fermion from a
fine-lattice to coarse-lattice, and derive a GWR with symmetry breaking
terms. Since the chiral limit can be taken independently of the
continuum limit, it is possible to obtain the PCAC relation on the
lattice. We also find an anomaly relation between the fermions on the
fine-lattice and the coarse-lattice.  We next show that the Dirac
operator satisfying the conventional GWR can be regarded as a mapping
from a torus to a sphere. Based on this identification, it can be
understood that a massless state corresponds to the South Pole of the
sphere and massive states (doublers) appear at the North Pole.  We want
to emphasize that massless and massive states can have different charges
under a L\"uscher's chiral transformation.  That is the reason why the
GW fermion escapes from the no-go theorem.  We discuss the BSTs of the
lattice fermions such as the Wilson fermion and the Domain Wall fermion.
It is shown that they must satisfy the GWR as approaching the
ultraviolet fixed point (FP). They share the following physical
properties near the continuum limit: (i) All doublers have the same mass
with order $O(a^{-1}$). This is compatible with the chiral symmetry
proposed by L\"uscher. (ii) Restoration of the ``rotational symmetry''.
So, the GW fermion has more symmetries than other lattice fermions.

  The present paper is organized as follows.  In section 2, we derive
 the GWR with a breaking term, chiral anomaly term and BST for Dirac
 operator.  Section 3 is devoted to properties of GWR.  We analyze
 block-spin transformed lattice fermion in section 4.  In section 5, we
 explicitly construct, for two dimensional case, the Dirac operator near
 the continuum limit using a conformal mapping.
 
\begin{flushleft}
 \large{{\bf 2.~GWR, Anomaly and BST }}
\end{flushleft}

In order to investigate the continuum limit which corresponds to a FP,
we will use the BST from a microscopic theory to a macroscopic
(effective) theory. Let $A_c$ be an action of the high frequency modes $
\psi_n$ and $\bar{\psi}_n$ with fine lattice index $n$, and $A$ be the
action of the low frequency modes$ \Psi_N$ and $\bar{\Psi}_N$ with
coarse-lattice index $N$. We assume that the action $A_c$ is bi-linear
in $ \psi_n$ and $\bar{\psi}_n$. The chiral transformation for the fine
lattice variables is given by

\begin{equation}
\delta \psi = -i\g_5\psi,
~~~\delta \bar{\psi} = \bar{\psi} (-i\g_5).
\label{eqn:chiral}
\end{equation}

\noindent
under which the action transforms as

\begin{equation}
\delta A_0(\psi, \bar{\psi}) =  A_0(\psi + \delta  \psi, \bar{\psi} + \delta  
\bar{\psi}) -A_0(\psi, \bar{\psi}).
\end{equation}

There are two kinds of chiral breaking for which $\delta A_0 \ne 0$: (i)
 The breaking due to the lattice regularization.  For example, the
 actions for the Wilson fermion and the Domain Wall fermion are not
 invariant under eq.(\ref{eqn:chiral}).  (ii) Explicit breaking such as
 the mass term.

We derive a general GWR for the coarse-lattice action $A$ in the
presence of the chiral breaking terms. The action is defined as

\begin{equation}
A \equiv - \ln \int_{\bar{\psi},\psi}    
K \exp( - A_0 [\psi, \bar{\psi}]) = \bar{\Psi}D \Psi .
\label{macro_a}
\end{equation}

\noindent
The block spin kernel, $K$, takes of the form 

\begin{equation}
K \sim \exp (- (\bar{\Psi} - \bar{\psi}f^*)\alpha (\Psi - f\psi)),
\end{equation}
\noindent
where the functions $f
(f^*)$ specify the BST and normalized as  $f_{Nn} f^*_{nM} = \delta_{NM}, ~~
f^*_{nN} f_{Nm} = \delta_{nm}$. 
We take ${\bf \alpha}_{MN} = \alpha \delta _{MN}$ so that $K$ becomes local.
The following formula for $K$ is useful:

\begin{eqnarray}
 \frac{\partial^l}{\partial \Psi} 
\alpha^{-1} \{\alpha,\gamma_5\}\alpha^{-1}
 \frac{\partial^l}{\partial \bar{\Psi}} K
 = (\bar{\Psi} - \bar{\psi}f^*)\{\alpha , \gamma_5\} (\Psi - f\psi)K.
\end{eqnarray}
We obtain the general GWR,

\begin{equation}
\bar{\Psi}(\gamma_5 D + D \gamma_5)\Psi = 
\bar{\Psi}D \alpha^{-1}\{\alpha ,\gamma_5\}
\alpha^{-1} D \Psi - i 
<\delta A_0>|_{\bar{\Psi}\Psi},
\label{eqn:GW1}
\end{equation}
\noindent
where $<\cdots>$ is defined as 

\begin{equation}
<\cdots> = \int_{\bar{\psi},\psi} \cdots 
 K \exp ( - A_0 [\psi, \bar{\psi}]) /
  \int_{\bar{\psi},\psi} K \exp ( - A_0 [\psi, \bar{\psi}]),
\end{equation}
\noindent and $|_{\bar{\Psi}\Psi}$ implies the component of
 $\bar{\Psi}\Psi$.  In section 4, we discuss the symmetry breaking term corresponding to the contribution $<\delta A_0>|_{\bar{\Psi}\Psi}$ for the Wilson fermions.

Next we consider a massive fermion on the fine lattice whose action
takes the bilinear form, $A_0 = \bar{\psi}(D_0 + m)\psi$.  Then, the
general GWR becomes
 
\begin{equation}
\bar{\Psi}(\gamma_5 D + D \gamma_5)\Psi = 
\bar{\Psi}D \alpha^{-1}\{\alpha ,\gamma_5\}
\alpha^{-1} D \Psi + 
2m \bar{\Psi}\alpha(\alpha + m + fD_0f^*)^{-1}
\gamma_5 (\alpha + m + fD_0f^*)^{-1}\alpha\Psi. 
\label{eqn:mgw}
\end{equation}
\noindent
Performing the Gaussian integral eq.(3), we find the relation between  the Dirac operator $D$
for the coarse-lattice variables and the $D_0$ for fine-lattice variables
 
\begin{eqnarray}
D^{-1} &=& \a^{-1} + f (D_{0} + m)^{-1}f^*.
\label{blockspin}
\end{eqnarray}

\noindent
This leads to the general GWR with the mass breaking term:

\begin{equation}
D\Gamma_5 + \Gamma_5D = 2m (1-\frac{D}{\alpha})\Gamma_5,
\label{eqn:pcac}
\end{equation}
where $ \Gamma_5 \equiv \g_5 (1-\frac{D}{\a})$.  It corresponds to the
``PCAC'' relation for the coarse-lattice fermions.  {}From this ``PCAC''
relation on the coarse-lattice, we can consider the chiral symmetry
breaking, e.g. the pion mass.  In eq.(\ref{eqn:pcac}), we observe that
the effective fermion mass on the coarse lattice should be regarded as
$m (1-\frac{D}{\alpha})$.  Note that the l.h.s. of eq.(\ref{eqn:pcac})
is written entirely with quantities defined for the coarse lattice,
while the r.h.s. contains also the microscopic information (microscopic
mass, $m$).

The general GWR eq.(\ref{eqn:pcac}) is obtained from the
eq.(\ref{eqn:mgw}) which is bi-linear in $\Psi$ and $\bar{\Psi}$. For
the field-independent relation, we obtain

\begin{equation}
\delta J  = - \frac{2}{\alpha}{\rm Tr}\gamma_5\{D- m(1-\frac{D}{\alpha})\}
\label{eqn:anomaly}
\end{equation}
\noindent 
where $\delta J$ means the shift of path integral measure under 
the chiral transformation eq.(\ref{eqn:chiral}) and the l.h.s corresponds to 
 chiral anomaly generated by fine-lattice fermion. 
 This  equality implies the anomaly generated  by microscopic fields 
 is saturated with  coarse-lattice (macroscopic) fields.
 Since $D$ includes a mass term, we have to eliminate the apparent effect
of the mass. 
 From $D-m(1-\frac{D}{\alpha}) \to D$, 
the apparent mass dependence of the l.h.s 
  is  vanished after redefinition of the Dirac operator $D$.

When $m=0$,
there is a remnant of chiral symmetry which is called L\"uscher's symmetry
\cite{lu},

\begin{equation}
\delta \Psi = -i\G_5\Psi,
~~~\delta \bar{\Psi} = \bar{\Psi} (-i\G_5).
\end{equation}
Eq.(\ref{eqn:anomaly}) implies the relation between microscopic and
macroscopic anomalies, since r.h.s of eq.(\ref{eqn:anomaly}) is ${\rm
Tr}\{2\Gamma_5\}$.  It is noted that the transformation of the low
frequency modes is given by \cite{iis}

\begin{equation}
\delta \Psi = <\delta \psi>,
~~~\delta \bar{\Psi} = <\delta \bar{\psi}>.
\end{equation}

\noindent
This gives another interpretation of the  L\"uscher's symmetry. 
Here we note that $<\psi>$ and $<\bar{\psi}>$ do not vanish
 because of the presence of the external fermions $\Psi$ and $\bar{\Psi}$.

In the proceeding  sections, we find the properties of fermions with GWR
  and carry out   BST explicitly for Wilson fermions.

\begin{flushleft}
 \large{{\bf 3.~GW fermion}}

\end{flushleft}

In this section we investigate  general properties of the free Dirac
operator satisfying the GWR. They are useful for discussing in 
section 4 that Dirac operators are gradually satisfying the GWR by 
BSTs,  and for constructing a Ginsparg-Wilson (GW) 
Dirac operator in section 5. 

\begin{flushleft}
 {\bf 3.1.~Properties of the GWR solutions}
\end{flushleft}
In a momentum space $\{p_{\mu}\},\mu=1\sim d$ $(d=even)$, the parity-even 
free GW Dirac operator can be written as 
\begin{eqnarray}
 D=i\gamma_{\mu}S_{\mu}(p)+A(p),  \label{eqn:1}
\end{eqnarray}
where $\gamma_{\mu}$ denote the hermitian gamma matrices, and
$S_{\mu}(p), A(p)$ are appropriate functions of momentum $p_{\mu}$.
Using the above expression, the GWR can be expressed as 
\begin{eqnarray}
 S_{\mu}S_{\mu}+\left(A-\frac{\alpha}{2}\right)^2=
\left(\frac{\alpha}{2}\right)^2.  \label{eqn:2}
\end{eqnarray}
It represents a d-dimensional
spherical surface $S^d$ with a radius of $\alpha /2$ in a
(d+1)-dimensional space $\{S_{\mu},A\}$. Since the momentum
space $\{p_{\mu}\}$ is equivalent to a d-dimensional torus $T^d$, we 
find that the GW Dirac operator is a mapping from $T^d$ to $S^d$:
\begin{eqnarray}
 D \,:\, T^d \longmapsto S^d.
\end{eqnarray}
It is noted that eq. (\ref{eqn:2}) has a rotational symmetry in a
subspace $\{S_{\mu}\}$. Although this is only a fake symmetry and not a
rotational one in the momentum space$\{p_{\mu}\}$, it is expected to
become a real rotational symmetry in the continuum limit.

Each point on the $S^d$ corresponds to each mode of the
propagator.  For $A=0$, a massless mode appears at the South Pole
(SP), and for $A=\alpha$, massive modes do at the North Pole (NP) in
Fig.$1$.  Both Poles are fixed points under the fake rotational
symmetry.  

\begin{figure}
\begin{center}
\epsfig{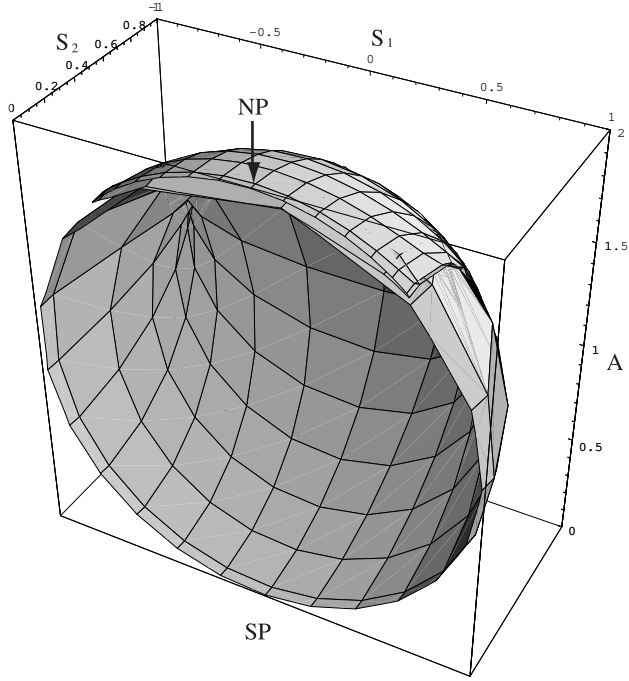}
\end{center}
\caption{A massless mode (SP) and massive mode (NP) in $\{ S_1 ,S_2 ,A \}$}
\end{figure}

As an example, consider the Dirac operator 
$D_{neu}(p)$ given by Neuberger \cite{neu},
\begin{eqnarray}
 D_{neu}(p)=i\gamma_{\mu}S_{\mu}^{neu}(p)+A_{\mu}^{neu}(p), \label{eqn:17}
\end{eqnarray}
\begin{center}
$\left\{
 \begin{array}{l}
   S_{\mu}^{neu}(p)=\sin p_{\mu} (H^2(p))^{-1/2}, \\
\ds{ A^{neu}(p)=1+(\sum_{\mu=1\sim d}(1-\cos p_{\mu})-M)(H^2(p))^{-1/2},} \\
\ds{H^2(p)  = \sum_{\nu=1\sim d}\sin^2 p_{\nu}+\left(\sum_{\nu=1\sim d}
    (1-\cos p_{\nu})-M\right)^2,}
 \end{array}
\right.$
\end{center}
\noindent
where $ 0<M<2$.
{}From a simple calculation, it is found that this Dirac operator satisfies
the equation of the d-dimensional sphere $S^d$,
$S_{\mu}^{neu}S_{\mu}^{neu}+(A^{neu}-1)^2=1$ with $\alpha=2$ and has the
rotational symmetry in a subspace $\{S_{\mu}^{neu}\}$. Fig.$1$ represents
this $S^{d=2}$ parameterized by momentum $p_{\mu}$.

The GW Dirac operator defines a smooth mapping from torus $T^d$ to
sphere $S^d$.  However, since they are different in topology, the
mapping is not one-to-one and several points on the torus may be mapped
to a point on the sphere.  For example, we observe in Fig.1 the North
Pole is realized for several values of momentum.  Fig.$2$ represents the
momentum dependence of $A(p)$, which shows the massless mode and the
degenerate massive modes.  From this figure we learn that the North Pole
is actually realized three times for this two dimensional example,
corresponding to the number of doublers.  In section 5, we will observe
the same feature in our new GW Dirac operator, different from the
Neuberger's one.

\begin{figure}
\begin{center}
\epsfig{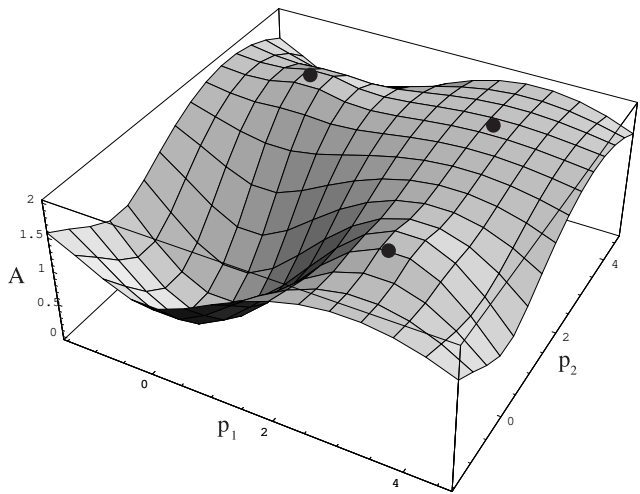}
\end{center}
\caption{Neuberger's Dirac operator in $\{ p_1 , p_2 , A \}$.  
Three doublers appear in association with three maxima indicated by dots.}
\end{figure}

Next, we consider the L$\ddot{\rm u}$scher's chiral transformation of the
massless mode $(S_{\mu}=0,A=0)$ and the massive modes $(S_{\mu}=0,A=\alpha)$.
Since the transformation is given by 
\begin{center}
 $ \left\{
 \begin{array}{lll}
 \psi & \rightarrow & \left[1 +i\theta\gamma_{5}
                      \left(1-\frac{D}{\alpha}\right)\right]\psi, \\
 \bar{\psi} & \rightarrow & 
   \bar{\psi}\left[1 +
    i\theta\left(1-\frac{D}{\alpha}\right)\gamma_{5}\right],
 \end{array}
 \right. $
\end{center}  
where $\theta$ is an infinitesimal parameter, one obtains for the massless
mode ($D=0$)
 \begin{center}
 $ \left\{
 \begin{array}{lll}
 \psi_{massless} & \rightarrow &\left(1+i\theta\gamma_{5}\right)
                   \psi_{massless}, \\
 \bar{\psi}_{massless} & \rightarrow & 
   \bar{\psi}_{massless}\left(1+i\theta\gamma_{5}\right),
  \end{array}
 \right. $
\end{center}
and for the massive modes ($D=\alpha$)
\begin{center}
 $ \left\{
 \begin{array}{lll}
 \psi_{massive} & \rightarrow & \psi_{massive}, \\
 \bar{\psi}_{massive} & \rightarrow & 
   \bar{\psi}_{massive}.
 \end{array}
 \right. $
\end{center}
It should be emphasized that the L\"uscher's chiral transformation
allows the mass for the doublers.  This fact looks incompatible with the
no-go theorem.  However, it is not the case, since the L\"uscher's
chiral symmetry does not satisfy one of the assumptions for the
no-go theorem, the locality of the chiral charge.

In the presence of interactions, the Dirac operator may not take the free
fields form $D=i\gamma_{\mu}S_{\mu}(p)+A(p)$, and cannot be interpreted
as a d-dimensional sphere $S^d$.  Nevertheless, the GW Dirac operator
always can be written as
\begin{equation}
 D=\frac{\alpha}{2}(1-U),
\end{equation}
where $U$ is a unitary operator. So the eigenvalues $\lambda$ are given
as 
\begin{equation}
 \lambda =\frac{\alpha}{2}(1-e^{i\theta}) \,,\,(-\pi\leq\theta<\pi).
\end{equation}
They distribute on a circle in a $\{Im(\lambda) ,Re(\lambda)\}$space
\cite{bh,bicudo,wb}.  For free field case, it is further possible to
consider the Dirac operator in terms of the modes of the propagator.

\begin{flushleft}
 {\bf 3.2.~The number of free parameters of the GW Dirac operator}
\end{flushleft}
Next, we will discuss the number of free parameters of the GW Dirac
operator. Let $D_{GW}$ and $D_{GW}^{\prime}$ be two GW Dirac
operators. In order for them to describe the same physics, they should
have the same ``fermion determinant'' and the same ``anomaly''. 
Introducing two hermite matrices, $H=\gamma_{5}D_{GW},
H^{\prime}=\gamma_{5}D_{GW}^{\prime}$, we may express the unitary
equivalence conditions as  

\begin{eqnarray}
 {\rm Det}\,H&=&{\rm Det}\, H^{\prime},\label{eqn:7} \\
 {\rm Tr}\,H&=&{\rm Tr}\,H^{\prime},\label{eqn:8} \\
 {\rm Tr}\,\left[H\right]^{n}&=&{\rm Tr}\,\left[H^{\prime}\right]^{n}   
  \,,\,\label{eqn:9}
\end{eqnarray}
where $2\leq n \leq N-1$ and $N$ denotes the total number of color,
flavor, spinor indices and lattice points. The equations (\ref{eqn:7})
and (\ref{eqn:8}) are the ``fermion determinant'' and the `` anomaly''
conditions.  Thus, the number of free parameters of the GW Dirac
operator is $N-2$, which is the number of
conditions of eq. (\ref{eqn:9}).  
This implies that there are many
solutions of the GWR.
Actually, in section 5 we will construct another
GW Dirac operator which differs from $D_{neu}$.

\begin{flushleft}
 \large{{\bf 4.~BST of Lattice Fermions }}
\end{flushleft}

In this section, we consider the Dirac operators which do not satisfy
the GWR, and show that the iterative use of BST makes them satisfy the
GWR approximately.  The Dirac operators gradually get the properties of
the GW fermion:

\begin{equation}
\left\{
 \begin{array}{ll}
  \rm (i) & \rm d-dimensional\,\, sphere\,{\it S^d} \,\,in\,\,the\,\, 
             {\it\{S_{\mu},A\}}space \\
  \rm (ii) &\rm rotational\,\, symmetry\,\, in\,\,the\,\, 
                   {\it\{S_{\mu}\}}subspace \\
  \rm (iii) &\rm  degenerate\,\, doubler\,\, masses.
 \end{array}
\right.
\end{equation}
For comparison, we also consider a BST of the GW fermion.

\begin{flushleft}
 {\bf 4.1.~Properties of the Wilson fermion}
\end{flushleft}
As a simple example, consider the 2D free Wilson Dirac operator
$D_{W}(p)$ given by  
\begin{eqnarray}
 D_{W}(p)=i\gamma_{\mu}S_{\mu}^{W}(p)+A^{W}(p), \label{eqn:11}
\end{eqnarray}
\begin{center}
 $\left\{
  \begin{array}{l}
    S_{\mu}^{W}(p)=\sin p_{\mu}, \\
    A^{W}(p)=\ds\sum_{\mu =1,2}(1-\cos p_{\mu}).
  \end{array}
 \right.$
\end{center}
The masses are defined by the values of $A(p)^W$
for $S_{\mu}^W(p)=0$. They are given in the case at hand by 
\begin{equation}
 A^{W}(p) = 
 \left\{
  \begin{array}{ll}
  0 &~{\rm at}~p_{1}=p_{2}=0 : {\rm original\,\, massless\,\, mode} \\
  2 &~{\rm at}~p_{1}=0, p_{2}=\pi~{\rm  or}~p_{1}=\pi,p_{2}=0 : {\rm doubler} \\
  4 &~{\rm at}~p_{1}=p_{2}=\pi  : {\rm doubler.}
  \end{array}
 \right.
\end{equation}
The doubler masses split into two values (Fig.$3$).
\begin{figure}
\begin{center}
\epsfig{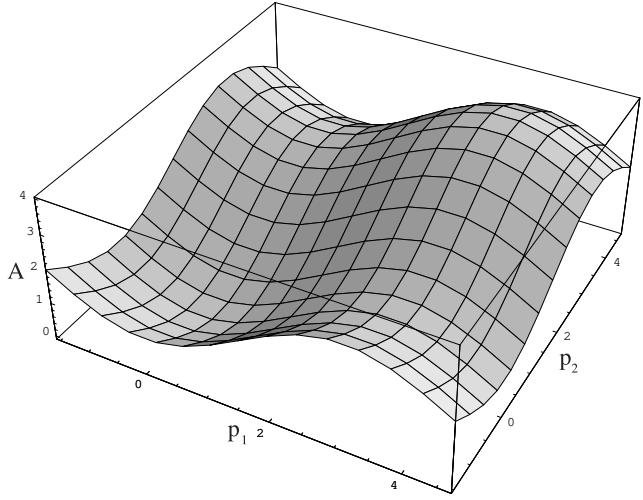}
\end{center}
\caption{Wilson-Dirac operator in $\{ p_1 , p_2 , A \}$}
\end{figure}
The Wilson fermion is
represented in the $\{S_{\mu}^W,A^W\}$space as Fig.$4$.  It 
is not a 2-dimensional sphere $S^2$, and there is no rotational symmetry in
the $\{S_{\mu}^W\}$space. 
\begin{figure}
\begin{center}
\epsfig{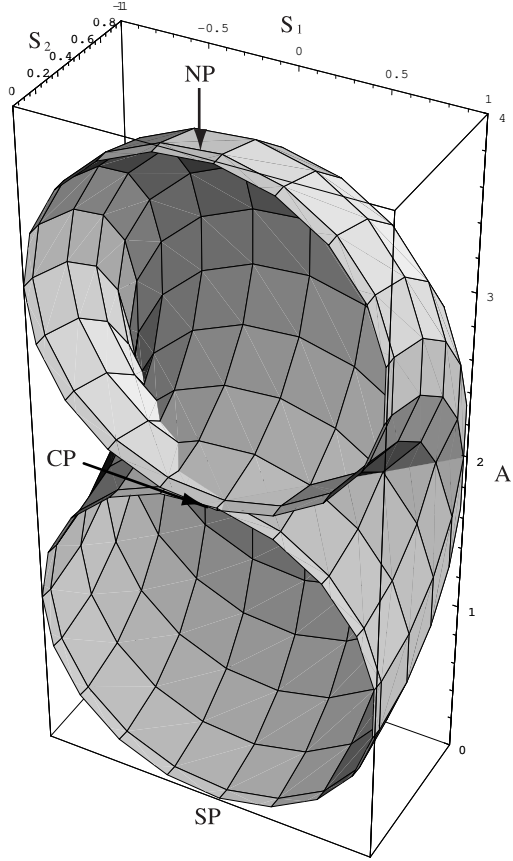}
\end{center}

\caption{Wilson fermion in $\{ S_1 ,S_2 ,A \}$}
\end{figure}
It follows from eq. (\ref{eqn:11}) 
\begin{equation}
 A^{W}=\sum_{\mu=1,2}\left(1 \pm\sqrt{1-(S_{\mu}^{W})^2}\right).
\end{equation}
In Fig.$4$ SP, central saddle point (CP), and NP correspond to the mass
 $A^W=$ 0, 2 and 4, respectively. 

\begin{flushleft}
 {\bf 4.2.~BST of the Wilson fermion}
\end{flushleft}
Next, we investigate that performing the BST eq.(\ref{blockspin}) 
of the Wilson fermion $(D^W=i\gamma_{\mu}S_{\mu}^{W}+A^{W})$, 
the fermion gets gradually the
properties of the GW fermion. By a special choice of $f_{Nn}$, 
 the BST can be defined as 
\begin{eqnarray}
 \left[\frac{-iS_{\mu}^{W}(p)\gamma_{\mu}+A^{W}(p)}
  {(S_{\mu}^{W}(p))^2+(A^{W}(p))^2}\right]_{N}
 &=&\frac{1}{2\alpha}
   +\frac{1}{2}\left[\frac{-iS_{\mu}^{W}(\frac{p}{2})\gamma_{\mu}
    +A^{W}(\frac{p}{2})}
  {(S_{\mu}^{W}(\frac{p}{2}))^2+(A^{W}(\frac{p}{2}))^2}\right]_{N-1},
  \label{eqn:BST}
\end{eqnarray}
where  $0 < \alpha < \infty$, the $N$ of $[ \cdots ]_{N}$ denotes the
N-th BST. Each BST can be constructed from eq.(\ref{blockspin})
 with a specific
choice of the function $f_{Nn}$.  From eq. (\ref{eqn:BST}) we can obtain  
\begin{eqnarray}
 \left[S_{\mu}^{W}(p)\right]_{N}&=&\left[\frac{X_{\mu}(\frac{p}{2})}
  {X_{\nu}^2(\frac{p}{2})+Y^2(\frac{p}{2})}\right]_{N-1},\label{eqn:14} \\
 \left[A^W(p)\right]_{N}&=&\left[\frac{Y(\frac{p}{2})}
  {X_{\nu}^2(\frac{p}{2})+Y^2(\frac{p}{2})}\right]_{N-1},\label{eqn:15}
\end{eqnarray}
where
\begin{center}
 $\left\{
  \begin{array}{lll}
   \left.X_{\mu}(\frac{p}{2})\right|_{N-1}&=&\ds\frac{1}{2}\left[
    \frac{S_{\mu}^{W}(\frac{p}{2})}
    {(S_{\nu}^{W}(\frac{p}{2}))^2+(A^{W}(\frac{p}{2}))^2}\right]_{N-1}, \\
   \left.Y(\frac{p}{2})\right|_{N-1}&=& \ds      
    \frac{1}{2\alpha}+\frac{1}{2}\left[\frac{A^{W}(\frac{p}{2})}
    {(S_{\nu}^{W}(\frac{p}{2}))^2+(A^{W}(\frac{p}{2}))^2}\right]_{N-1},
  \end{array}
 \right.$ 
\end{center}
then we can discuss the block-spin transformed Dirac operator in
the $\{S_{\mu}^W,A^W\}$space which represented as Fig.$5$. 
\begin{figure}
\begin{center}
\epsfig{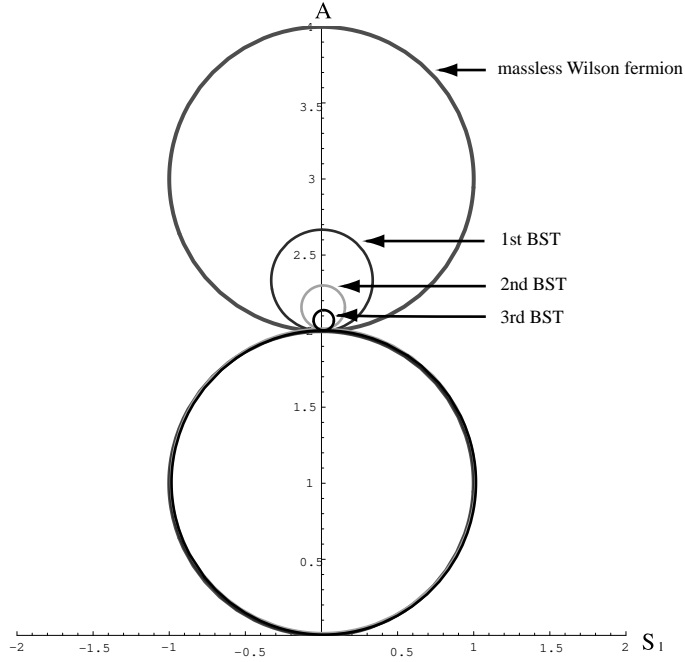}
\end{center}
\caption{BST of massless Wilson fermion. $\alpha$ is set to 2.}
\end{figure}
{}From this, we can visually understand that the more the Wilson
fermion is repeatedly transformed by the BST, the more it accurately
satisfies the GW fermion properties. Now verify analytically
them. Considering the GWR,  
\begin{eqnarray}
 \frac{A(p)}{(S_{\nu}(p))^2+(A(p))^2} 
 &=&\frac{1}{\alpha} ,\label{eqn:gwr}
\end{eqnarray}
and using eqs. (\ref{eqn:14}) and (\ref{eqn:15}), we get
\begin{eqnarray}
 \left[\frac{A^{W}(p)}
 {(S_{\nu}^{W}(p))^2+(A^{W}(p))^2}\right]_{N} 
 &=&\frac{1}{\alpha}+\left(\frac{1}{2}\right)^{N}
    \left\{\left[\frac{A^{W}(\frac{p}{N})}
    {(S_{\nu}^{W}(\frac{p}{N}))^2+(A^{W}(\frac{p}{N}))^2}
     \right]_{0}-\frac{1}{2\alpha}\right\}\nonumber \\
 && \nonumber \\
 &\longrightarrow & \frac{1}{\alpha}\,,\,(N\rightarrow\infty).
 \label{eqn:18}
\end{eqnarray}
Thus it is found that the Wilson fermion (and also other fermions)
satisfies the GWR in FP $(N\rightarrow\infty)$. Let us consider the
degeneracy of the doubler masses. {}From eqs. (\ref{eqn:14}) and
(\ref{eqn:15}) the masses are given as
\begin{eqnarray}
  \left[A^W\right]_{N} = \left\{ 
    \begin{array}{ll}
  0,& \\
  \ds\left.\frac{1}{Y(\frac{p^{\prime}}{2})}\right|_{N-1} 
     &= \ds\frac{2\alpha\left[A^W(\frac{p^{\prime}}{2})\right]_{N-1}}
  {\alpha+\left[A^W(\frac{p\prime}{2})\right]_{N-1}}, \\
  \ds\left.\frac{1}{Y(\frac{p^{\prime\prime}}{2})}\right|_{N-1}
     &= \ds\frac{2\alpha\left[A^W(\frac{p^{\prime\prime}}{2})\right]_{N-1}}
  {\alpha+\left[A^W(\frac{p^{\prime\prime}}{2})\right]_{N-1}},
     \end{array}
    \right. 
 \label{eqn:20}
\end{eqnarray}
where
$p_{\mu}^{\prime}=(0,N\pi)\,,\,p_{\mu}^{\prime\prime}=(N\pi,N\pi)$, 
fold following inequality, 
\begin{equation}
   0\,<\,
   \left[A^W(\frac{p^{\prime}}{2})\right]_{N-1}\,<\,
   \left[A^W(\frac{p^{\prime\prime}}{2})\right]_{N-1},
\end{equation}
and the ratio of the doubler masses $(\neq 0)$ is 
\begin{eqnarray}
 \left[\frac{A^W(p^{\prime\prime})}{A^W(p^{\prime})}\right]_{N}
  &=&
 \frac
  {\left[A^W(\frac{p^{\prime\prime}}{2})\right]_{N-1}+\alpha
  \left[\frac{A^W(\frac{p^{\prime\prime}}{2})}
  {A^W(\frac{p^{\prime}}{2})}\right]_{N-1}}
  {\left[A^W(\frac{p^{\prime\prime}}{2})\right]_{N-1}+\alpha} \nonumber \\
 & & \nonumber \\
 & \longrightarrow & 1 \,,\,(N\rightarrow\infty),
\end{eqnarray}
then at the FP the doubler masses ($\neq 0$) become degenerate. 
On the other hand, from eq. (\ref{eqn:20}) 
we can obtain the mass correction $\delta^{(N)}m$ by the N-th BST,
\begin{eqnarray}
  [A^{W}]_{N} = [A^{W}]_{N-1} + \delta^{(N)}m ,
\end{eqnarray} 
\begin{eqnarray}
  \delta^{(N)}m = \left[A^{W}\frac{\alpha-A^{W}}
                          {\alpha+A^{W}}\right]_{N-1}. \label{eqn:mc}
\end{eqnarray}
This implies that the doubler masses $(\neq 0)$ approaches to $\alpha$
by the BST, and the massless mode is a fixed point of the BST. 
Therefore, as far as $\alpha$ is finite ($\alpha<\infty$),
all doubler masses become the same mass $\alpha$ as $N\rightarrow\infty$. 
Thus it is understood that performing the BST on the fermion, the
fermion gradually gets the properties of the GW fermion.

\begin{flushleft}
 {\bf 4.3.~BST of the GW fermion}
\end{flushleft}
For comparison, consider the BST of the GW fermion. 
The Neuberger Dirac operator in eq. (\ref{eqn:17}) satisfies the GWR, 
\begin{eqnarray}
 \left. S_{\mu}^{neu}(p)S_{\mu}^{neu}(p)
 +\left(A^{neu}(p)-\frac{\alpha}{2}\right)^2
 =\left(\frac{\alpha}{2}\right)^2\right|_{micro}.
\end{eqnarray}
Thus, the BST is given as
\begin{eqnarray}
 \left[S_{\mu}^{neu}(p)\right]_{MACRO}&=&
  \left.\frac{2\alpha \,A^{neu}(\frac{p}{2})\,S_{\mu}^{neu}(\frac{p}{2})}
  {\left[S_{\nu}^{neu}(\frac{p}{2})\right]^2
  +4\left[A^{neu}(\frac{p}{2})\right]^2}\right|_{micro}, \\
 \left[A^{neu}(p)\right]_{MACRO}&=&\left.
  \frac{4\alpha\left[A^{neu}(\frac{p}{2})\right]^2}
  {\left[S_{\mu}^{neu}(\frac{p}{2})\right]^2
  +4\left[A^{neu}(\frac{p}{2})\right]^2}\right|_{micro},
\end{eqnarray}
and one can show that the
$\left[S_{\mu}^{neu}(p)\right]_{MACRO}$ and the $[A^{neu}(p)]_{MACRO}$
satisfy the GWR also, 
\begin{eqnarray}
 \left. S_{\mu}^{neu}(p)S_{\mu}^{neu}(p)
 +\left(A^{neu}(p)-\frac{\alpha}{2}\right)^2
 =\left(\frac{\alpha}{2}\right)^2\right|_{MACRO}.
\end{eqnarray}
After all we have obtained the structure of chiral symmetry near the 
FP. That is, the GW fermion flows into the FP keeping the GWR. The
other fermions flow into the FP getting the properties of the GW fermion 
approximately, and satisfy the GWR at FP at the end. 
Thus fermions have at least the approximate GWR near FP. 

In the above discussion, we have considered only about the massless fermion
case which had a bare mass $m_{0} = 0$.
Now, consider a massive fermion case with a non-zero bare mass
$(m_{0}\neq0)$. From the fact that the all modes are massive $(\neq 0)$
and that the mass correction by the BST is given from eq. (\ref{eqn:mc}),
the mass  will
become  $\alpha$ at the FP (Fig.$6$).
\begin{figure}
\begin{center}
\epsfig{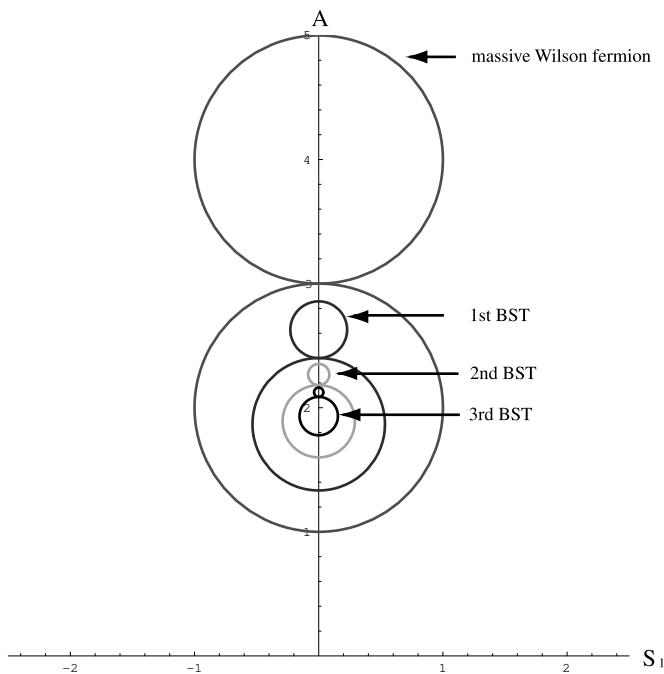}
\end{center}
\caption{BST of massive Wilson fermion. $\alpha$ is set to 2.}
\end{figure}
And the Dirac operator will satisfy the GWR.\footnote{
Chandrasekharan introduced a bare mass for the GWR, but 
his GWR is different from ours\cite{sc}.}

\begin{flushleft}
{\bf 4.4.~The domain wall fermion and massless states}
\end{flushleft}

We consider here fermion masses 
in the domain wall formulation\cite{kaplan,shamir,fs}. 
The domain wall fermion can be interpreted as  
a multi-flavor Wilson fermion with a  negative mass.
The number of flavor corresponds to the fifth dimensional size. Its rule
 can be explained in view of the BST or the renormalization group as
 follows: increasing the flavor number or the fifth dimensional size in
 the domain wall fermion corresponds to looking at the behavior of the
 4-dimensional fermion at larger scale, which would be described
 by an iteration of the BST's.  In the infinite size limit, it is known
 that the 4-dimensional effective fermion satisfies the GWR\cite{kn}.
 It just corresponds to the continuum limit.  The degeneracy  of the doublers
 mass and the restoration of the ``rotational invariance'' are
 essentially the same as those of one flavor Wilson fermion near the
 continuum limit.

In this subsection, we concentrate on 
 the massless state in the domain wall fermion 
 related with the the fifth dimensional size.

Let $N_{S}$ be the fifth dimensional size for the domain wall fermions
\cite{kaplan,shamir,fs}. 
Decompose a Dirac spinor $\Psi(x_{\mu},x_{5})$ as 
\begin{eqnarray}
 \Psi(x_{\mu},x_{5}) = P_{L}\Psi_{L}(x_{\mu},x_{5}) 
                      + P_{R}\Psi_{R}(x_{\mu},x_{5}),
\end{eqnarray}
where $P_{L,R}=(1\pm\gamma_{5})/2$. We impose the following
appropriate boundary condition for $x_{5}$ on $\Psi(x_{\mu},x_{5})$,
\begin{eqnarray}
 \Psi_{L}(x_{\mu},x_{5}+N_{S}) = \Psi_{R}(x_{\mu},x_{5}), \\ 
 \Psi_{R}(x_{\mu},x_{5}+N_{S}) = \Psi_{L}(x_{\mu},x_{5}).
\end{eqnarray}
The domain wall Dirac operator $D_{DW}(p_{\mu},p_{5})$ in a
momentum space $\{p_{\mu},p_{5}\}$ can be written as
\begin{eqnarray}
 D_{DW}(p_{\mu},p_{5}) = i \sum_{\mu =1\sim 4}\gamma_{\mu}S_{\mu}^{DW}
                                (p_{\mu})
                   +A^{DW}(p_{\mu},p_{5}), 
\end{eqnarray}
where 
\begin{center}
 $\left\{
  \begin{array}{l}
    S_{\mu}^{DW}(p_{\mu})=\sin p_{\mu}, \\
    A^{DW}(p_{\mu},p_{5})= - M +(1-\cos p_{5} +i\gamma_{5}\sin p_{5}) 
                     +{\displaystyle\sum_{\mu =1\sim 4}}(1-\cos p_{\mu}),
  \end{array}
 \right.$
\end{center}
and $0<M<2$ , $p_{5}= \pi n/N_{S},\,\,n=1\sim 2N_{S}$.
The lowest mass $m_{eff}(p_{5})$ which depends on $p_{5}$ is given by
\begin{eqnarray}
    m_{eff}(p_{5})=A^{DW}(p_{\mu}=0,p_{5})
           = - M +(1-\cos p_{5} +i\gamma_{5}\sin p_{5}). 
\end{eqnarray}
For each case of $\gamma_{5}=\pm 1$, we denote each mass as 
\begin{eqnarray}
    m_{eff}^{\pm}(p_{5})= - M +(1-\cos p_{5} \pm i\sin p_{5}).
\end{eqnarray}
This yields a constraint,
\begin{eqnarray}
  -M \leq Re\{m_{eff}^{\pm}(p_{5})\} \leq 2-M. 
\end{eqnarray}
In the limit $N_{S} \rightarrow\infty$, each
$Re\{m_{eff}^{\pm}(p_{5})\}$ has a continuum spectrum with two zeros due
to the periodicity of $p_{5}$.  Then the number of total zeros are four.
According to the sign of each imaginary part $Im\{m_{eff}^{\pm}\}$, only
half of four zeros are allowed by causality.  Therefore two allowed
zeros are regarded as massless modes corresponding to $\gamma_{5}=\pm 1$, 
and we obtain a massless fermion theory.  For a finite
$N_{S}$, each $Re\{m_{eff}^{\pm}(p_{5})\}$ has a discrete spectrum. Thus
$Re\{m_{eff}^{\pm}(p_{5})\}$ does not have zero in general, and 
we obtain in this case a massive theory.\footnote[3]
{ It is possible, however, to make
$Re\{m_{eff}^{\pm}(p_{5})\}=0$ for a certain value $p_{5}$ by making a
fine-tuning for  $M$. Thus we may 
construct a fine-tuned massless fermion theory for finite $N_{S}$.}
As $N_{S}$ increases, the minimum value of $|Re\{m_{eff}^{\pm}\}|$
decreases, and finally becomes zero which corresponds to a massless mode.

The zero has no mass correction via the BST, that is a fixed point of the
BST. Thus we can obtain a massless fermion theory which satisfies the
GWR at the FP of the BST.

\begin{flushleft}
\large{\bf 5.~Lattice Fermions near the Continuum Limit}
\end{flushleft}

In this section, we show that the Neuberger's Dirac operator is not the
unique GW Dirac operator, by constructing another free GW Dirac
operator, $D_{GW}=i\gamma_{\mu}S_{\mu}+A$, for two dimensional case.  As
is shown in eq.(16), the construction of $D_{GW}$ is equivallent to find
a mapping from a 2D torus $T^2$ to a 2D sphere $S^2$ in the
$(S_1,S_2,A)$ space,
\begin{eqnarray}
 D_{GW}(p) & : & T^2 \longmapsto S^2 .
\end{eqnarray}
We obtain this mapping by using a combination of conformal mappings.
Before going into the details of the construction, let us explain our
idea.  We divide the torus equally into four regions (see {\bf I}
below).  Take one of the regions and map it to the Sourthern hemisphere
conformally. It is important that this conformal mapping is unique
due to the Riemann's theorem.  Each of three other regions is similarly
mapped conformally to the Northern hemisphere.  Our mapping is
constructed by continuously connecting the four conformal mappings.

We will construct such a mapping with the following steps.
\begin{description}
\item[I :] 
  Divide the momentum space, a complex plane $P = p_1 +i p_2$, into 
four domains 	   
 \begin{equation}
 \begin{array}{cccccc}
  P^{(1)}&=&\{(p_1,p_2)|&-\frac{\pi}{2}\leq p_1 \leq\frac{\pi}{2}&,&
          -\frac{\pi}{2}\leq p_2 \leq\frac{\pi}{2}\},\\
  P^{(2)}&=&\{(p_1,p_2)|&\frac{\pi}{2}\leq p_1 \leq\frac{3\pi}{2}&,&
         -\frac{\pi}{2}\leq p_2 \leq\frac{\pi}{2}\},\\
  P^{(3)}&=&\{(p_1,p_2)|&-\frac{\pi}{2}\leq p_1 \leq\frac{\pi}{2}&,&
         \frac{\pi}{2}\leq p_2 \leq\frac{3\pi}{2}\},\\
  P^{(4)}&=&\{(p_1,p_2)|&\frac{\pi}{2}\leq p_1 \leq\frac{3\pi}{2}&,&
           \frac{\pi}{2}\leq p_2 \leq\frac{3\pi}{2}\}.
 \end{array} \label{eqn:26}
 \end{equation}
 It is arranged in such a way that $(p_1,p_2)=(0,0)$ in
 $P^{(1)}$ and  $(p_1,p_2)=(\pi,0),(0,\pi),(\pi,\pi)$ in
 $P^{(2),(3),(4)}$ correspond to a massless mode and massive modes,
 respectively.
\item[II :]
 Regarding the $(S_1,S_2)$ space as a complex plane with $S=S_1 +i S_2$
and $\bar{S}=S_1 -i S_2$, the GWR is written as 	   
 \begin{equation}
  S \bar{S} + \left(A-\frac{\alpha}{2}\right)^2 =
  \left(\frac{\alpha}{2}\right)^2 .
 \end{equation}
 This gives 
 \begin{equation}
  A_{\pm}=\frac{\alpha}{2}\pm \sqrt{\left(\frac{\alpha}{2}\right)^2 
  -S\bar{S}}, \label{eqn:28}
 \end{equation}
where $A_{-}$($A_{+}$) corresponds to the Southern (Northern) Hemisphere. It contains
a massless (massive) mode as the South (North) Pole of $S^2.$ We
	   construct conformal mappings $S^{(1),(2),(3),(4)}$ from
	   $P^{(1),(2),(3),(4)}$ to disks with a radius of
	   $\alpha/2$,\footnote[5]{In order for the four points,
	   $(p_1,p_2)=(0,0),(\pi,0),(0,\pi),(\pi,\pi)$ in
	   $P^{(1),(2),(3),(4)}$, to behave as poles, we require
	   $S^{(1),(2),(3),(4)}=0$ at these points.  This is indeed the
	   case thanks to the Riemann's mapping theorem which guarantees
	   that $S_{\mu}\sim ip_{\mu}$ for $|p_{\mu}| << 1$.}
\begin{equation}
 S^{(i)}(P^{(i)})=\left\{(S_1,S_2)\left|\right.
  S\bar{S}=S_1S_1+S_2S_2\leq \left(
  \frac{\alpha}{2}\right)^2\right\}\,,\,i=1\sim 4.\label{eqn:29}
\end{equation}
\item[III :]
 From eqs. (\ref{eqn:28}) and (\ref{eqn:29}) we can get $A_{-}$ and $A_{+}$
 corresponding to $P^{(1)}$ and $P^{(2),(3),(4)}$, respectively,
 as explicit function of momenta $p_{\mu}$.
\end{description}
In this way we can construct a solution of the GWR, 
 $D_{GW}=i\gamma_{\mu}S_{\mu}+A$. 

First let us construct the disk $S^{(1)}$ as the conformal mapping,
$P^{(1)} \mapsto V^{(1)} \mapsto S^{(1)}$, where $V^{(1)}$ denotes 
upper half plane mapped from $P^{(1)}$. The conformal mapping from
$P^{(1)}$ to $V^{(1)}$ is defined as 
\begin{eqnarray}
 V^{(1)}={\rm sn}\left(P^{(1)}+\frac{i}{2}K^{\prime},k\right),\label{eqn:30}
\end{eqnarray}
where sn is an elliptic function with modulus $k$, and $K^{\prime}$ is
the complete elliptic integral of the first kind with complementary
modulus $k^{\prime}=\sqrt{1-k^2}$. The elliptic function sn$(P,k)$ is a
periodic function of $4K\,,\,2iK^{\prime}$. In our case
$K=\frac{\pi}{2}\,,\,K^{\prime}=\pi$, and $k=(\sqrt{2}-1)^2$.  Also
considering the conformal mapping from $V^{(1)}$ to $S^{(1)}$, we 
obtain the mapping $P^{(1)} \mapsto S^{(1)}$,
\begin{eqnarray}
 S^{(1)}&=&\frac{V^{(1)}-i(\sqrt{2}+1)}
                {V^{(1)}+i(\sqrt{2}+1)}
                \cdot i\frac{\alpha}{2} \nonumber \\             
                                      &&\nonumber \\   
        &=&\frac{{\rm sn}\left(P^{(1)}+\frac{i}{2}K^{\prime},k\right)
               -i(\sqrt{2}+1)}
              {{\rm sn}\left(P^{(1)}+\frac{i}{2}K^{\prime},k\right)
               +i(\sqrt{2}+1)}\cdot i\frac{\alpha}{2}. \label{eqn:31}
\end{eqnarray}
Similarly, we construct the mapping, 
$P^{(2),(3),(4)} \mapsto S^{(2),(3),(4)}$, paying attention to
continuity at boundaries.
The mapping, $P^{(4)} \mapsto V^{(4)}$ is given by 
\begin{eqnarray}
 V^{(4)}&=&
   {\rm sn}\left(-(P^{(4)}-2K-iK^{\prime})
    +\frac{i}{2}K^{\prime},k\right) \nonumber \\
       &=&
   {\rm sn}\left(-(P^{(4)}-iK^{\prime})
    +\frac{i}{2}K^{\prime},k\right) \nonumber \\
       &=&
   -{\rm sn}\left(-(P^{(4)}+\frac{i}{2}K^{\prime})
     +2iK^{\prime},k\right) \nonumber \\
       &=&
   {\rm sn}\left(P^{(4)}+\frac{i}{2}K^{\prime},k\right),
\end{eqnarray}
where we used the properties ${\rm sn}(P+2K,k)=-{\rm sn}(P,k)$\,,\,
${\rm sn}(P+2iK^{\prime},k)={\rm sn}(P,k)$\,,\,
${\rm sn}(-P,k)=-{\rm sn}(P,k)$.
Thus we obtain the mapping, $P^{(4)} \mapsto S^{(4)}$ 
\begin{eqnarray}
 S^{(4)}&=&\frac{{\rm sn}\left(P^{(4)}+\frac{i}{2}K^{\prime},k\right)
               -i(\sqrt{2}+1)}
              {{\rm sn}\left(P^{(4)}+\frac{i}{2}K^{\prime},k\right)
               +i(\sqrt{2}+1)}\cdot i\frac{\alpha}{2}. \label{eqn:33}
\end{eqnarray}
As for $ V^{(2),(3)}$, it is convenient to take the complex conjugate 
$\overline{P}$.  The mappings,  
$\overline{P^{(2),(3)}}\,\mapsto \,V^{(2),(3)}\,
\mapsto \, S^{(2),(3)}$ are given by
\begin{eqnarray}
 V^{(2)}&=&
   {\rm sn}\left(-\overline{(P^{(2)}-2K)}
    +\frac{i}{2}K^{\prime},k\right) \nonumber \\
       &=&
   -{\rm sn}\left(-\overline{P^{(2)}}
    +\frac{i}{2}K^{\prime},k\right) \nonumber \\
       &=&
   {\rm sn}\left(\overline{(P^{(2)}
    +\frac{i}{2}K^{\prime})},k\right) \nonumber \\
       &=&
   \overline{\rm sn}\left(P^{(2)}
    +\frac{i}{2}K^{\prime},k\right),
\end{eqnarray}
and
\begin{eqnarray}
 V^{(3)}&=&
   {\rm sn}\left(\overline{(P^{(3)}
    -iK^{\prime})}+\frac{i}{2}K^{\prime},k\right) \nonumber \\
       &=&
   {\rm sn}\left(\overline{(P^{(3)}+\frac{i}{2}K^{\prime})}
     +2iK^{\prime},k\right) \nonumber \\
  &=&\overline{\rm sn}\left(P^{(3)}
        +\frac{i}{2}K^{\prime},k\right).
\end{eqnarray}
We find that 
\begin{eqnarray}
 S^{(2),(3)}&=&\frac{\overline{\rm sn}\left(P^{(2),(3)}
                +\frac{i}{2}K^{\prime},k\right)-i(\sqrt{2}+1)}
              {\overline{\rm sn}\left(P^{(2),(3)}
                +\frac{i}{2}K^{\prime},k\right)
                +i(\sqrt{2}+1)}\cdot i\frac{\alpha}{2}. \label{eqn:36}
\end{eqnarray}
Using eqs. (\ref{eqn:28}),(\ref{eqn:31}),(\ref{eqn:33}),(\ref{eqn:36}),
and $P=P_1+iP_2\,,\,S=S_1+iS_2$, we obtain a solution of the 2D free GW
Dirac operator, $D_{GW}=i\gamma_{\mu}S_{\mu}+A$.
 For the momentum domains $P^{(1),(4)}$ ,
 \begin{eqnarray}
   S_1 &=&2Re\left({\rm sn}\left(P
    +\frac{i}{2}K^{\prime},k\right)\right)\frac{\alpha\,f_{+}}{2}, \\
   S_2 &=&\left\{(\sqrt{2}-1){\rm sn}\left(P
    +\frac{i}{2}K^{\prime},k\right)\overline{\rm sn}\left(P
    +\frac{i}{2}K^{\prime},k\right)
    +(\sqrt{2}+1)\right\}\frac{\alpha\,f_{+}}{2}, \nonumber\\
    && \\
   A &=&\frac{\alpha}{2}\left\{1\mp
    \sqrt{4Im\left({\rm sn}\left(P
    +\frac{i}{2}K^{\prime},k\right)\right)f_{+}}\right\},\label{eqn:39}
 \end{eqnarray}
 where $-$ and $+$ in eq. (\ref{eqn:39}) correspond to $P^{(1)}$ and 
 $P^{(4)}$, respectively, and $f_{\pm}^{-1}$ are defined as
 \begin{eqnarray}
   (f_{\pm})^{-1} &=&(\sqrt{2}-1){\rm sn}\left(P
    +\frac{i}{2}K^{\prime},k\right)\overline{\rm sn}\left(P
    +\frac{i}{2}K^{\prime},k\right) \nonumber \\
    &&+(\sqrt{2}+1)\pm 2Im\left({\rm sn}\left(P
    +\frac{i}{2}K^{\prime},k\right)\right).
 \end{eqnarray}
 For the momentum domains $ P^{(2),(3)}$,
  \begin{eqnarray}
   S_1 &=&2Re\left({\rm sn}\left(P
    +\frac{i}{2}K^{\prime},k\right)\right)\frac{\alpha\,f_{-}}{2}, \\
   S_2 &=&\left\{(\sqrt{2}-1){\rm sn}\left(P
    +\frac{i}{2}K^{\prime},k\right)\overline{\rm sn}\left(P
    +\frac{i}{2}K^{\prime},k\right)
    -(\sqrt{2}+1)\right\}\frac{\alpha\,f_{-}}{2} ,\nonumber\\
    && \\
   A &=&\frac{\alpha}{2}\left\{1+
    \sqrt{4Im\left({\rm sn}\left(P
    +\frac{i}{2}K^{\prime},k\right)\right)f_{-}}\right\}.\label{eqn:43}
  \end{eqnarray}
In Fig.$7$, we show this solution in the $\{S_{1},S_{2},A\}$
space. Although singular points appear on the equator,\footnote{In
Fig.7, the white stripe around the equator is generated by a bug of the
graphical softwere we use. After analytical calculations, it can be
shown that upper hemisphere and lower hemisphere are continuously
connected with each other.}  the Dirac operator
$D_{GW}$ is a smooth function of $p_{\mu}$.

\begin{figure}
\begin{center}
\epsfig{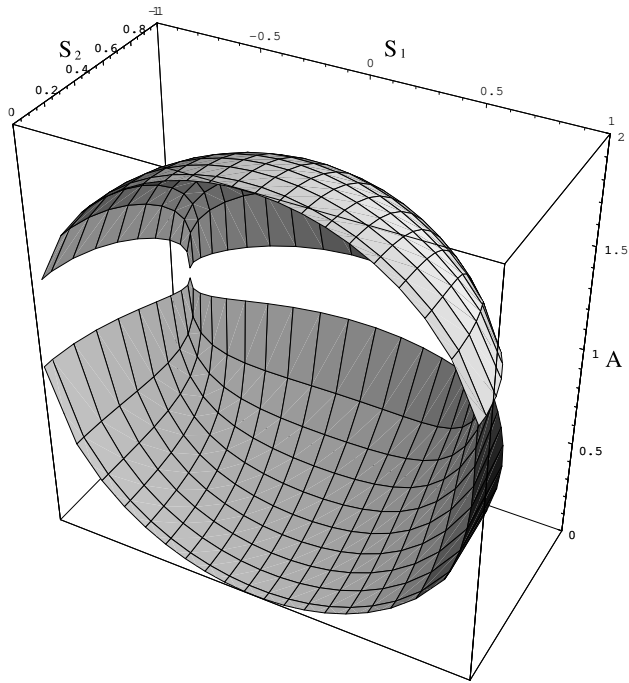}
\end{center}
\caption{Dirac operator using elliptic function in $\{ S_1 ,S_2 ,A\}$}
\end{figure}

\begin{flushleft}
 \large{{\bf 6.~Summary and Discussions }}
\end{flushleft}
We have investigated behaviors of lattice Dirac operators near the fixed
 point (the continuum limit).  For the purpose of our approach, we used 
 a bock-spin transformation method.  The analysis lead us to the chiral
 anomaly and the GWR with chiral breaking terms. The breaking measures
 the distance between the fixed point action and the general lattice
 fermions.  It was found that features of the operators are (i) masses of
 doublers are completely degenerate (ii) ``rotational'' symmetry.  When
 we start from a Wilson fermion with a chiral breaking, it acquires these two
 properties after a few block-spin steps.
 
 In the case of multi-flavor Wilson fermions (the domain wall fermion), 
  we found massless modes   as  a fifth dimension size becomes infinite.
 Even if we adopt more general lattice fermion, we could find massless modes
 because the  fifth dimension size essentially  
controls the macroscopic fermion mass as the BST does.  

In the 2-dimensional case, we constructed a Dirac operator,
 in additon to Neuberger's one,
 which satisfies the GWR using elliptic functions.
 This example  teaches us the unique existence 
 of the fixed point action (chiral symmetric lattice fermion),
 if it is possible to exchange  between a massless mode and one of doublers 
 because of uniqueness  for conformal mapping by the Riemann's theorem.

 We are grateful to Y. Igarashi and K. Itoh for reading our manuscript 
 carefully and invaluable comments. 
H.S. is supported in part by the Grants-in-Aid for Scientific Research 
 No. 12640259 from Japan Society for the Promotion of Science.


\begin{thebibliography}{99}
\bibitem{lu} M.~L{\"u}scher, Phys.\ Lett. {\bf B428} (1998) 342.
\bibitem{gw} P.~Ginsparg and K.~Wilson, Phys.\ Rev.\ {\bf D25} (1982) 2649.
\bibitem{lu1}  M.~L{\"u}scher, Nucl.\ Phys. {\bf B538} (1999) 515;
Nucl.\ Phys. {\bf B549} (1999) 295;  Nucl.\ Phys. {\bf B568} (2000) 162.


\bibitem{su} H.~So and N.~Ukita, Phys.\ Lett. {\bf B457} (1999) 314. 
\bibitem{bie} W.~Bietenholz, Mod.\ Phys.\ Lett. {\bf A14} (1999) 51.
\bibitem{ak}  T.~Aoyama, Y.~Kikukawa, Phys.\ Rev. {\bf D59} (1999) 054507.
\bibitem{hn}  H.~Neuberger, Phys.\ Rev.\ Lett. {\bf 81} (1998) 4060.
\bibitem{hjlu} P.~Hernandez, K.~Jansen and M.~Luscher, Nucl. \ Phys. {\bf B552}
 (1999) 363.
\bibitem{hs0}  H.~Suzuki, Prog.\ Theor.\ Phys. {\bf 101} (1999) 1147.
\bibitem{kf0} K.~Fujikawa, Phys.\ Rev. {\bf D60} (1999) 074505.
\bibitem{ab} A.~Borici,  hep-lat/0001019. 
\bibitem{chiu}  T.~Chiu, Phys.\ Lett. {\bf B467} (1999) 112.
\bibitem{hn1} H.~Neuberger, hep-lat/9910040. 
\bibitem{ukqcd} UKQCD Collaboration, hep-lat/9912033. 
\bibitem{cg} C.~Gattringer, hep-lat/0003005.
\bibitem{hhh} P.~Hasenfratz, S.~Hauswirth, K.~Holland, T.~Jorg, F.~Niedermayer and  U.~Wenger, hep-lat/0003013. 
\bibitem{cgih} C.~ Gattringer and I.~Hip,  Phys.\ Lett. {\bf B480} (2000) 112.

\bibitem{hf} P.~Hasenfratz, hep-lat/9709110.

\bibitem{iis} Y.~Igarashi, K.~Itoh and  H.~So, Phys.\ Lett.{\bf B479} (2000) 336; hep-th/0006180. 
 


\bibitem{neu} H.~Neuberger, Phys.\ Lett. {\bf B417} (1998) 141;  Phys.\ Lett. {\bf B427} (1998) 353.


\bibitem{bh} W.~Bietenholz and I.~Hip,  Nucl.\ Phys. {\bf B570} (2000) 423.
\bibitem{bicudo} P.~J.~de~A~Bicudo, hep-lat/9912015. 
\bibitem{wb} W.~Bietenholz,  hep-lat/0007017.


\bibitem{sc} S.~Chandrasekharan, Phys.\ Rev. {\bf D60} (1999) 074503.

\bibitem{kaplan} D.~Kaplan, Phys.\ Lett. {\bf B288} (1992) 342.
\bibitem{shamir} Y.~Shamir, Nucl.\ Phys. {\bf B406} (1993) 90.
\bibitem{fs}  V.~Furman and Y.~Shamir, Nucl.\ Phys. {\bf B439} (1995) 265. 
\bibitem{kn}  Y.~Kikukawa and T.~Noguchi,  hep-lat/9902022. 

\end{thebibliography}
\end{document}